\documentclass[amsmath,showpacs,nofootinbib,12pt]{revtex4-2}
\usepackage{graphicx}
\usepackage{dcolumn}
\usepackage{bm}
\usepackage{color} 
\usepackage{slashed}
\begin{document}
\newcommand{\hs}{\hspace*{0.5cm}}
\newcommand{\vs}{\vspace*{0.5cm}}
\newcommand{\be}{\begin{equation}}
\newcommand{\ee}{\end{equation}}
\newcommand{\bea}{\begin{eqnarray}}
\newcommand{\eea}{\end{eqnarray}}
\newcommand{\ben}{\begin{enumerate}}
\newcommand{\een}{\end{enumerate}}
\newcommand{\bde}{\begin{widetext}}
\newcommand{\ede}{\end{widetext}}
\newcommand{\nn}{\nonumber}
\newcommand{\crn}{\nonumber \\}
\newcommand{\Tr}{\mathrm{Tr}}
\newcommand{\non}{\nonumber}
\newcommand{\noi}{\noindent}
\newcommand{\al}{\alpha}
\newcommand{\la}{\lambda}
\newcommand{\bet}{\beta}
\newcommand{\ga}{\gamma}
\newcommand{\va}{\varphi}
\newcommand{\om}{\omega}
\newcommand{\pa}{\partial}
\newcommand{\+}{\dagger}
\newcommand{\fr}{\frac}
\newcommand{\bc}{\begin{center}}
\newcommand{\ec}{\end{center}}
\newcommand{\Ga}{\Gamma}
\newcommand{\de}{\delta}
\newcommand{\De}{\Delta}
\newcommand{\ep}{\epsilon}
\newcommand{\varep}{\varepsilon}
\newcommand{\ka}{\kappa}
\newcommand{\La}{\Lambda}
\newcommand{\si}{\sigma}
\newcommand{\Si}{\Sigma}
\newcommand{\ta}{\tau}
\newcommand{\up}{\upsilon}
\newcommand{\Up}{\Upsilon}
\newcommand{\ze}{\zeta}
\newcommand{\ps}{\psi}
\newcommand{\Ps}{\Psi}
\newcommand{\ph}{\phi}
\newcommand{\vph}{\varphi}
\newcommand{\Ph}{\Phi}
\newcommand{\Om}{\Omega}
\newcommand{\AdrHEPC}{Phenikaa Institute for Advanced Study and Faculty of Basic Science, Phenikaa University, Yen Nghia, Ha Dong, Hanoi 100000, Vietnam}

\title{Abelian charge inspired by family number} 

\author{Phung Van Dong} 
\email{Corresponding author; dong.phungvan@phenikaa-uni.edu.vn}
\author{Tran Ngoc Hung}
\email{hung.tranngoc@phenikaa-uni.edu.vn}
\author{Duong Van Loi}
\email{loi.duongvan@phenikaa-uni.edu.vn}
\affiliation{\AdrHEPC} 
\date{\today}

\begin{abstract}

Quark has an electric charge either $-1/3$ or $2/3$ and a baryon number $1/3$, where the divisions $3$'s match the color number. Although the electric charge and the baryon number have a nature distinct from the color charge, the matching is necessary for the standard model or a relevant $B-L$ extension consistent at quantum level, since the relevant anomaly $[SU(2)_L]^2U(1)_A$ for $A=Y$ or $B-L$ must vanish. If elementary particles have a new $U(1)$ charge differently from $A$, such anomaly is not cancelled for each family. However, if we demand that the anomaly is cancelled over all families, this relates the color number to the family number instead of the electric charge and baryon number, and interestingly the family number guides us to a novel $U(1)$ theory. We will discuss the implication of this theory for neutrino mass, recent $W$-boson mass anomaly, FCNC, and particle colliders.            

\end{abstract}

\maketitle

\section{Motivation}
The weak isospin symmetry $SU(2)_L$, which forms isodoublets $l_L=(\nu_L,e_L)$ and $q_L=(u_L,d_L)$ while the right-handed partners $e_R$, $u_R$, and $d_R$ are isosinglets, is not completed. The electric charge $Q$, which is $\mathrm{diag}(0,-1)$ for $l_L$ and $\mathrm{diag}(2/3,-1/3)$ for $q_L$, is neither commuted nor closed algebraically with $SU(2)_L$, because of $[Q,T_1\pm i T_2]=\pm (T_1\pm i T_2)\neq 0$ and $\mathrm{Tr}(Q)\neq 0$. Since $[T_3,T_1\pm i T_2]=\pm (T_1\pm i T_2)$ is similar to $Q$, a new abelian charge $Y\equiv Q-T_3$ arises, such that $[Y,T_1\pm i T_2]=0=[Y,T_3]$. This defines a complete group, $SU(2)_L\otimes U(1)_Y$, which establishes the standard electroweak model \cite{g,w,as}. The model successfully works at quantum level because every anomaly vanishes for which the most important one is $[SU(2)_L]^2 U(1)_Y\sim 3 Y_{q_L}+Y_{l_L}=3\times \fr 1 6 - \fr 1 2 =0$. This cancellation occurring for each family is due to the fact that the color number $3$ relates the electric charge of quark to that of lepton. Besides the hypercharge $Y$, the model contains an accident gauge symmetry of baryon number minus lepton number $B-L$, if a right-handed neutrino $\nu_R$ is simply imposed for each family. In this case, every anomaly is cancelled too for which the most important one is $[SU(2)_L]^2 U(1)_{B-L}\sim 3 [B-L]_{q_L}+[B-L]_{l_L}=3\times \fr 1 3 - 1 =0$. This cancellation is due to the fact that the color number $3$ relates the $B-L$ number of quark to that of lepton, analogous to the case of electric charge. Who arranges such matchings, since the electric charge and the $B-L$ number have a nature distinct from the color number? 

This work does not directly answer such question. We guess that at high energy the partial interactions are unified in a simple group, well-known as grand unification. The unification group would contain the color charge and weak isospin, besides neutral generators that determine the hypercharge and even $B-L$. This work does not search for a specific unification group. Instead, we investigate possible neutral charges that lead to the hypercharge. The hypothetical symmetry of $B-L$ although mentioned will not be addressed. Obviously, an individual neutral charge, called $X$, does not necessarily have the above property as of the hypercharge, i.e. $3 X_{q_L} + X_{l_L}\neq 0$ for a single family. The anomaly $[SU(2)_L]^2 U(1)_X$ is cancelled, needed to a correlative contribution with exotic particles or between families. The latter case would make a constraint on the number of families. The minimal grand unifications such as $SU(5)$ \cite{su5}, $SO(10)$ \cite{so10}, and Pati-Salam model \cite{ps} do not contain $X$, since they leave only the hypercharge and/or $B-L$, $T_{3R}$. The trinification \cite{trinification} and $E_6$ \cite{e6} as well as their variants \cite{hd,331} would provide such a $X$. 

In what follows, we search for such neutral charges and propose the model of interest. We discuss the new physics insights that the new proposal reveals. Interestingly, this approach provides a potential explanation to neutrino mass, $W$-boson mass deviation, compelling phenomena for flavor physics, as well as hints for particle colliders.             

\section{Proposal of the model}

Instead of $Y,B-L$ charges and even their combination, we look for an entirely alternative charge $X$ such that the relation $3 X_{q_L} + X_{l_L}=0$ does not hold for each family. That said, the new electroweak group takes the form, 
\be SU(2)_L\otimes U(1)_X\otimes U(1)_N,\label{dttntn02}\ee where $3 X_{q_L}+X_{l_L}\neq 0$ for each family, and the charge $N$ is necessarily included such that $U(1)_X\otimes U(1)_N$ is broken down to $U(1)_Y$ at high energy, i.e. $N=Y-X$. The $N$ charge obeys $3N_{q_L}+N_{l_L}=-(3X_{q_L}+X_{l_L})\neq 0$ as $X$ does, for each family. Thus $X,N$ have a nature different from $Y$ (and $B-L$ too), but the breaking of $X,N$ explains $Y$ and the matching. 

Given a grand unification, $X$ and $N$ would be unified with $SU(2)_L$ in a higher isospin group. Hence, the magnitudes of $X,N$  can be fixed as the neutral charges of the large group. Obviously, $X$ (or $N$) is not originated from a charge, such as $T_{3R}$ or $B-L$, as in minimal left-right symmetry, Pati-Salam group, and $SO(10)$. Looking for trinification, we find that $T_{8L,R}$ have a property of $X$, namely $X\sim T_{8L}+T_{8R}$, while $N$ is followed by $N=Y-X$. Note that fermion doublets may be arranged in trinification triplets/antitriplets. Hence, we restrict ourselves by assigning $X_{l_L}=x$ for all lepton doublets, $X_{q_L}=x$ for a number of quark doublets, while $X_{q_L}=-x$ for the rest of quark doublets. Here the normalization of $x$ is left arbitrarily. The condition $3 X_{q_L} + X_{l_L}\neq 0$ for each family simply requires $x\neq 0$. Summing over $N_f$ families, the anomaly $[SU(2)_L]^2 U(1)_{X}\sim 3 (m -n)x + N_f x$ vanishes if $N_f=3(n-m)$, where $m$ and $n=N_f-m$ denote numbers of $q_L$ doublets that possess $x$ and $-x$, respectively. It is clear that the similar anomaly for $N$, i.e. $[SU(2)_L]^2 U(1)_N$, vanishes as a result. Hence, the family number $N_f=3(n-m)$ is an integer multiple of the color number, $3$, for which fixing $N_f=3$ leads to $n=2$ and $m=1$. We define two kinds of family indices, like $a,b,...$ run over 1,2,3 according to $N_f=3$ and $\al,\beta,...$ run over 1,2 according to $n=2$. We then assign $(X,N)$ charges to fermion representations as
in Table \ref{tab1}, where the color and isospin quantum numbers are also included for completeness.
\begin{table}
\bc
\begin{tabular}{lcccc}
\hline\hline 
Field & $SU(3)_C$ & $SU(2)_L$ & $U(1)_X$ & $U(1)_N$\\
\hline 
$l_{aL}=\begin{pmatrix}
\nu_{aL}\\
e_{aL}\end{pmatrix}$ & 1 & 2 & $x$ & $-1/2-x$ \\
$\nu_{aR}$ & 1 & 1 & $x$ & $-x$\\
$e_{aR}$ & 1 & 1 & $x$ & $-1-x$\\
$q_{\al L}=\begin{pmatrix}
u_{\al L}\\
d_{\al L}\end{pmatrix}$ & 3 & 2 & $-x$ & $1/6+x$ \\
$u_{\al R}$ & 3 & 1 & $-x$ & $2/3+x$\\
$d_{\al R}$ & 3 & 1 & $-x$ & $-1/3+x$\\
$q_{3 L}=\begin{pmatrix}
u_{3 L}\\
d_{3 L}\end{pmatrix}$ & 3 & 2 & $x$ & $1/6-x$ \\
$u_{3 R}$ & 3 & 1 & $x$ & $2/3-x$\\ 
$d_{3R}$ & 3 & 1 & $x$ & $-1/3-x$\\
$H=\begin{pmatrix}
H^+_1\\
H^0_2
\end{pmatrix}$ & 1 & 2 & 0 & $1/2$\\
$\Phi=\begin{pmatrix}
\Phi^+_1\\
\Phi^0_2
\end{pmatrix}$ & 1 & 2 & $-2x$ & $1/2+2x$\\
$\chi$ & 1 & 1 & $-2x$ & $2x$\\
\hline\hline
\end{tabular}
\caption[]{\label{tab1} Field representation under the gauge symmetry.}
\ec
\end{table}     
             
That said, the third quark doublet has a $X$-charge as of lepton doublets, while the first and second quark doublets have an opposite $X$-charge. Hence, $3X_{q_L}+X_{l_L}=4x$ for the third family, while $3X_{q_L}+X_{l_L}=-2x$ for the first and second families. The anomaly $[SU(2)_L]^2 U(1)_X\sim 3X_{q_L}+X_{l_L}$ vanishes when summing over three families. It is verified that all the other anomalies, say $[SU(3)_C]^2U(1)_X$, $[\mathrm{gravity}]^2 U(1)_X$, those for $X\rightarrow N$, $[U(1)_X]^2 U(1)_N$, $U(1)_X [U(1)_N]^2$, $[U(1)_X]^3$, and $[U(1)_N]^3$, vanish too. The anomaly cancellation is due to the fact that the family number matches the color number. Last, but not least, the right-handed neutrinos are presented, required for anomaly cancellation. Additionally, for symmetry breaking and mass generation, we have introduced the scalar fields: (i) $H$ is identical to that of the standard model, (ii) $\Phi$ couples the two kinds of quarks $\bar{q}_{\al}$ and $q_3$ necessary for recovering CKM matrix, and (iii) $\chi$ couples to right-handed neutrinos $\nu_R \nu_R$ responsible for neutrino mass generation.  

\begin{figure}[h]
\bc
\includegraphics[scale=0.9]{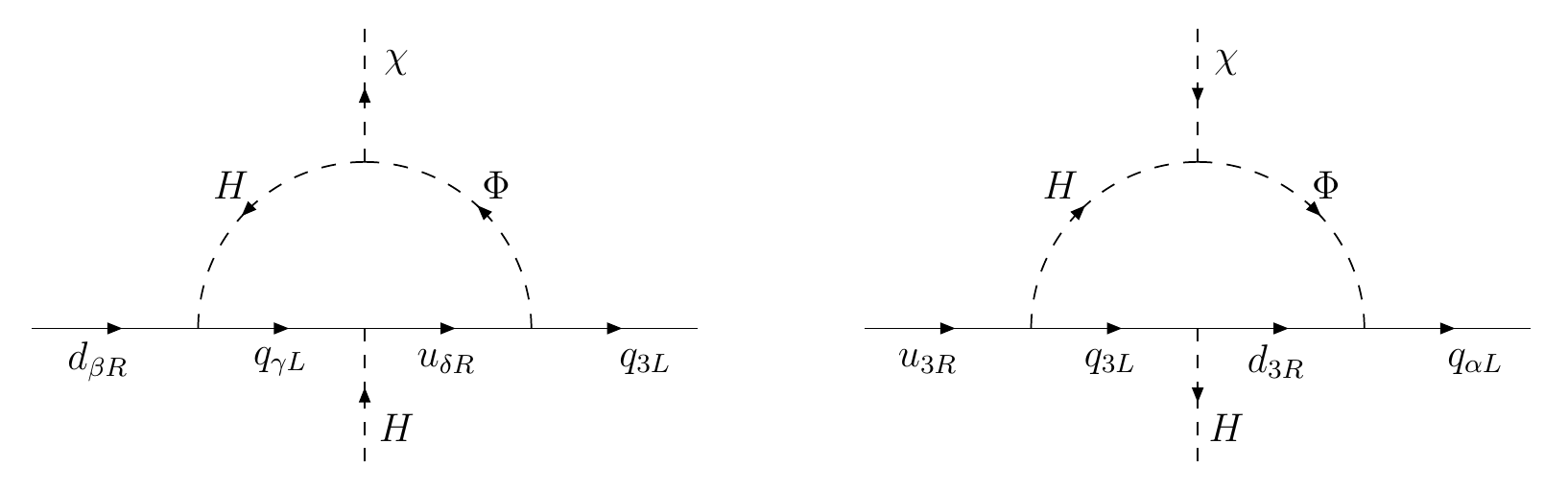}
\caption[]{\label{figadd}  Yukawa couplings of $\Phi'$ and $\tilde{\Phi}'$ types generated by corresponding Feynman diagrams (left and right) that identify $\Phi'\sim H\chi^*$ and $\tilde{\Phi}'\sim \tilde{H}\chi$, respectively.}
\ec
\end{figure}

Up to the gauge fixing and ghost terms, the total Lagrangian takes the form, 
\be \mathcal{L}=\mathcal{L}_{\mathrm{kin}}+\mathcal{L}_{\mathrm{Yuk}}-V.\ee The kinetic term is 
\bea \mathcal{L}_{\mathrm{kin}} &=&\sum_F \bar{F} i\ga^\mu D_\mu F + \sum_S (D^\mu S)^\dagger (D_\mu S)\crn
&&-\fr 1 4 G_{p\mu\nu} G^{\mu\nu}_p -\fr 1 4 A_{j\mu\nu}A^{\mu\nu}_j -\fr 1 4 B_{\mu \nu} B^{\mu\nu} -\fr 1 4 C_{\mu\nu} C^{\mu\nu},\eea where $F$ and $S$ run over fermion and scalar multiplets, respectively. The covariant derivative and field strength tensors are given by  
\bea && D_\mu = \pa_\mu + i g_s t_p G_{p\mu} + i g T_j A_{j\mu}+ i g_X X B_\mu + i g_N N C_\mu,\crn
&& G_{p\mu\nu}=\pa_\mu G_{p\nu}-\pa_\nu G_{p \mu} - g_s f_{pqr} G_{q\mu}G_{r\nu},\crn
&& A_{j\mu\nu}=\pa_\mu A_{j\nu}-\pa_\nu A_{j \mu} - g \ep_{jkl} A_{k\mu}A_{l\nu},\crn
&& B_{\mu\nu}=\pa_\mu B_\nu - \pa_\nu B_\mu,\hs C_{\mu\nu}=\pa_\mu C_\nu -\pa_\nu C_\mu,\nn \eea where $(g_s,g,g_X, g_N)$, $(t_p, T_j, X, N)$, and $(G_{p\mu}, A_{j\mu}, B_\mu, C_\mu)$ denote the coupling constants, the generators, and the gauge bosons of the $(SU(3)_C, SU(2)_L, U(1)_X, U(1)_N)$ groups, respectively. A kinetic mixing term between the two $U(1)$ gauge fields is suppressed, due to its small effect.  The Yukawa Lagrangian and the scalar potential are given by 
\bea \mathcal{L}_{\mathrm{Yuk}} &=& h^e_{ab}\bar{l}_{aL} H e_{bR}+h^\nu_{ab} \bar{l}_{aL} \tilde{H} \nu_{bR}+\fr 1 2 f^\nu_{ab} \bar{\nu}^c_{a R} \nu_{bR}\chi\crn
&&+ h^d_{\al \beta} \bar{q}_{\al L} H d_{\beta R}+ h^u_{\al \beta} \bar{q}_{\al L} \tilde{H} u_{\beta R} +h^d_{33}\bar{q}_{3L} H d_{3R}+ h^u_{33} \bar{q}_{3L} \tilde{H} u_{3R} 
\crn
&&+ h'^d_{\al 3} \bar{q}_{\al L} \Phi d_{3 R} + h'^u_{3\beta } \bar{q}_{3L} \tilde{\Phi} u_{\beta R} +h'^d_{3\beta }\bar{q}_{3L} \Phi' d_{\beta R} + h'^u_{\al 3} \bar{q}_{\al L} \tilde{\Phi}' u_{3 R} \crn
&&+H.c. \label{dttndt01}
\eea
\bea
V &=& \mu^2_1 H^\dagger H + \mu^2_2 \Phi^\dagger \Phi + \mu^2_3 \chi^\dagger \chi +\mu_4 [(\Phi^\dagger H) \chi+H.c.] \crn
&&+ \la_1(H^\dagger H)^2+\la_2(\Phi^\dagger \Phi)^2+\la_3 (\chi^\dagger \chi)^2\crn
&& +\la_4 (H^\dagger H)(\chi^\dagger \chi) +\la_5 (\Phi^\dagger \Phi) (\chi^\dagger \chi)\crn
&&+ \la_6 (H^\dagger H)(\Phi^\dagger \Phi) + \la_7(H^\dagger \Phi) (\Phi^\dagger H). \eea   
Above, $f^\nu$, $h$'s, and $\la$'s are dimensionless, while $\mu_{1,2,3,4}$ have a mass dimension. Additionally, we define $\tilde{H}=i\sigma_2 H^*$ and $\tilde{\Phi}=i \sigma_2 \Phi^*$. The last two terms in (\ref{dttndt01}) before the $H.c.$ are radiatively induced by one-loop diagrams despited in Fig. \ref{figadd} respectively, through $\Phi,H$ exchanges and their couplings to quarks, and that the scalar trilinear coupling $\Phi^\dagger H \chi$ is set by $\mu_4$. Computing these diagrams, we identify $\Phi'\equiv H\chi^*(\mu_4/16\pi^2M^2)$ and $\tilde{\Phi}'\equiv \tilde{H}\chi (\mu_4/16\pi^2M^2)$, where $1/16\pi^2$ is the loop factor, and $M$ is the characteristic large mass of heavy scalar field running in the loop; additionally, the Yukawa couplings induced are $h'^d_{3\beta}\equiv h'^u_{3\delta}h^{*u}_{\ga \delta}h^d_{\ga \beta}$ and $h'^u_{\al 3}\equiv h'^d_{\al 3}h^{*d}_{33}h^u_{33}$. For convenience, in the following we will match $\langle \Phi'\rangle =\langle H\rangle \langle \chi\rangle^* (\mu_4/16\pi^2M^2)$ with $\langle \Phi\rangle$ by rescaling $h'^d_{3 \beta}$ and $h'^u_{\al 3}$ appropriately. 

At this point, there are two extra models alternative to the current model, given in order
\ben \item A generic model may be built by introducing two new fundamental Higgs doublets in addition to the usual $H$, i.e. $\Phi\sim (2,-2x,1/2+2x)$ and $\Phi'\sim (2,2x,1/2-2x)$, which transform under the symmetry (\ref{dttntn02}) as $H\chi$ and $H\chi^*$, respectively. This model is not significantly different from the current model for phenomenological aspects interested in this work, thus not considered. 
\item A minimal model can be built by suppressing both $\Phi$ and $\Phi'$. Instead, their Yukawa interactions in (\ref{dttndt01}) are given by corresponding effective interactions by substituting $\Phi=H\chi/M'$ and $\Phi'=H\chi^*/M'$, where $M'$ is a cutoff scale set by a more fundamental theory. Interestingly, this model does not encounter any tree-level FCNC coupled to scalars in contrast to the above models. However, it presents the phenomenological aspects interested in this work similar to the current model, thus skipped too. \een    

The gauge symmetry is broken as follows 
\bc \begin{tabular}{c} 
$SU(3)_C\otimes SU(2)_L \otimes U(1)_X\otimes U(1)_N$ \\
$\downarrow\La$\\
$SU(3)_C\otimes SU(2)_L \otimes U(1)_Y$ \\
$\downarrow v_{1,2}$\\
$SU(3)_C\otimes U(1)_Q$ \\
\end{tabular}\ec Here the vacuum expectation values (VEVs) are given by 
\bea &&\langle \chi\rangle =\La/\sqrt{2},\\ 
&& \langle H\rangle = (0,v_1/\sqrt{2}),\\ 
&&\langle \Phi \rangle =(0,v_2/\sqrt{2}),\eea and we require $\La\gg v_{1,2}$ and $v^2_1+v^2_2=(246\ \mathrm{GeV})^2$ for consistency with the standard model. As stated, the combined field $\Phi'$ gets a value matching that of $\Phi$, i.e. $\langle \Phi'\rangle = (0,v_2/\sqrt{2})$.

\section{Fermion mass}

Substituting the VEVs to the Yukawa Lagrangian, we obtain fermion masses. First, charged leptons gain a mass via $h^e$ coupling to be \be [m_e]_{ab}=-h^e_{ab}\fr{v_1}{\sqrt{2}}.\ee Diagonalizing this mass matrix, we get the masses of electron, muon, and tau, as usual. 

The neutrinos $\nu_{L,R}$ gain a Dirac mass via $h^\nu$ coupling, 
\be [m_D]_{ab}=-h^\nu_{ab} \fr{v_1}{\sqrt{2}},\ee while $\nu_R$ obtains a Majorana mass via $f^\nu$ coupling,
\be [m_M]_{ab}=-f^\nu_{ab}\fr{\La}{\sqrt{2}}.\ee The total mass matrix of neutrinos takes the form, 
\bea \mathcal{L}_{\mathrm{Yuk}}\supset -\fr 1 2 (\bar{\nu}_{aL}, \bar{\nu}^c_{aR})
\begin{pmatrix}
0 & [m_D]_{ab}\\
[m^T_D]_{ab} & [m_M]_{ab}\end{pmatrix}
\begin{pmatrix}
\nu^c_{bL}\\
\nu_{bR}\end{pmatrix}+H.c.\nn\eea Because of $v_1\ll \La$, i.e. $m_D\ll m_M$, the active neutrinos $\sim \nu_L$ receive a small mass via the canonical seesaw,
\be [m_\nu]_{ab}= -[m_D m^{-1}_M m^T_D]_{ab} \sim v^2_1/\La.\ee Taking $v_1\sim 100$ GeV and $\La\sim 5$--10 TeV, the observed values $m_\nu\sim 0.1$ eV require $h^\nu\sim 10^{-5}$, given that $f^\nu\sim 1$. The heavy neutrinos $\sim \nu_R$ obtain a mass $m_M$ at $\La$ scale.

Concerning quarks, their masses are given by 
\be [m_d]_{\al \beta} =-h^d_{\al \beta}\fr{v_1}{\sqrt{2}},\hs [m_d]_{33}=-h^d_{33}\fr{v_1}{\sqrt{2}},\ee \be
[m_d]_{\al 3} = -h'^d_{\al 3}\fr{v_2}{\sqrt{2}},\hs [m_d]_{3\beta}=-h'^d_{3\beta}\fr{v_2}{\sqrt{2}},\ee for down-type quarks, while \be [m_u]_{\al \beta} =-h^u_{\al \beta}\fr{v_1}{\sqrt{2}},\hs [m_u]_{33}=-h^u_{33}\fr{v_1}{\sqrt{2}},\ee \be
[m_u]_{\al 3} = -h'^u_{\al 3}\fr{v_2}{\sqrt{2}},\hs [m_u]_{3\beta}=-h'^u_{3\beta}\fr{v_2}{\sqrt{2}},\ee for up-type quarks. Hence, the quark masses are governed by two weak scales $v_{1,2}$ as well as two kinds of couplings $h$ and $h'$. The small mixing between the third quark family and the first two quark families  can be understood by $h'< h$ and $v_2<v_1$. Of course, diagonalizing the quark mass matrices, we get the masses of $u,d,c,s,t,b$ and CKM matrix, as expected.  

\section{Gauge boson mass}

Substituting the VEVs to the scalar kinetic term, i.e. $\sum_S (D^\mu S)^\dagger (D_\mu S)$, we get gauge boson masses. First notice that the gluons are massless since $SU(3)_C$ is conserved, unbroken. Concerning the electroweak sector, the charged gauge boson $W^\pm=(A_1\mp i A_2)/\sqrt{2}$ is a physical field by itself with mass, 
\be m^2_W=\fr{g^2}{4}(v^2_1+v^2_2).\ee Whereas, the neutral gauge bosons $A_3$, $B$, and $C$ mix via a mass matrix, 
\be \mathcal{L}_{\mathrm{kin}}\supset \fr 1 2 (A_3, B, C)M^2
\begin{pmatrix}
A_3\\
B\\
C
\end{pmatrix},\ee where
\be M^2=\begin{pmatrix}
\fr{g^2}{4}(v^2_1+v^2_2) & x g g_X v^2_2& -\fr{gg_N}{4}[(1+4x)v^2_2+v^2_1]\\
x g g_X v^2_2 & 4x^2 g^2_X (\La^2+v^2_2) & -g_X g_N[x(1+4x) v^2_2+4x^2\La^2] \\
-\fr{gg_N}{4}[(1+4x)v^2_2+v^2_1] & -g_X g_N[x(1+4x) v^2_2+4x^2\La^2] & g^2_N[4x^2 \La^2+\fr{v^2_2}{4}(1+4x)^2+\fr{v^2_1}{4}]
\end{pmatrix}.\ee
 
The mass matrix of neutral gauge bosons has an exact, zero eigenvalue (i.e. photon mass) corresponding to the exact mass eigenstate (i.e. photon field), such as  
\be \fr{A}{e}=\fr{A_3}{g}+\fr{B}{g_X}+\fr{C}{g_N},\ee which is normalized by \be \fr{1}{e^2}=\fr{1}{g^2}+\fr{1}{g^2_X}+\fr{1}{g^2_N}.\ee The photon field properly couples to the electric charge $Q=T_3+Y=T_3+X+N$, since the above solution can be obtained by substituting each generator by corresponding gauge field over coupling, say $Q$ by $A/e$, $T_3$ by $A_3/g$, $X$ by $B/g_X$, and $N$ by $C/g_N$. The electromagnetic coupling is identified as $e=gs_W$ through the sine of the Weinberg's angle, while the tan of this angle is given by $t_W=g_Y/g$, where the hypercharge coupling is obtained by $g_Y=g_X g_N/\sqrt{g^2_X+g^2_N}=g_X s_\theta =g_N c_\theta$, thus $t_\theta = g_N/g_X$. We rewrite 
\be A= s_W A_3 + c_W (s_\theta B + c_\theta C). \ee We choose two other fields: the usual $Z$ field, \be Z=c_W A_3 - s_W (s_\theta B + c_\theta C),\ee which is given orthogonally to $A$, as usual, and the new field, 
\be Z'=c_\theta B -s_\theta C,\ee which is given orthogonally to the hypercharge field in parentheses. 

In the new basis $(A,Z,Z')$, the photon $A$ is decoupled as a physical field, while the remainders $Z,Z'$ mix through a mass matrix,
\be \mathcal{L}_{\mathrm{kin}}\supset \fr 1 2 (Z,Z')
\begin{pmatrix}
m^2_Z & m^2_{ZZ'}\\
m^2_{ZZ'} & m^2_{Z'}
\end{pmatrix}
\begin{pmatrix}
Z\\
Z'
\end{pmatrix},\ee where 
\bea m^2_{Z} &=& \fr{g^2}{4c^2_W}(v^2_1+v^2_2),\\
m^2_{ZZ'}&=&\fr{g^2t_W}{2s_{2\theta}c_W}\left[s^2_{\theta} v^2_1+(s^2_\theta+4x)v^2_2\right],\\
m^2_{Z'}&=& \fr{g^2t^2_W}{s^2_{2\theta}}\left[s^4_\theta v^2_1+(s^2_\theta +4x)^2v^2_2+16x^2 \La^2\right].\eea  Diagonalizing the $Z,Z'$ mass matrix, we obtain two physical fields,
\bea Z_1 &=& c_\varphi Z - s_\varphi Z',\\
Z_2 &=& s_\varphi Z + c_\varphi Z',\eea where the $Z$-$Z'$ mixing angle ($\varphi$) obeys 
\be t_{2\varphi}=\fr{2m^2_{ZZ'}}{m^2_{Z'}-m^2_{Z}}\simeq \fr{s_{2\theta}\left[s^2_\theta v^2_1+(s^2_\theta +4x)v^2_2\right]}{16s_W x^2 \La^2},\ee and the $Z_{1,2}$ masses are given by
\bea m^2_{Z_1} &=& \fr 1 2 \left[m^2_{Z}+m^2_{Z'}-\sqrt{(m^2_Z-m^2_{Z'})^2+4m^4_{ZZ'}}\right]\simeq m^2_Z-\fr{m^4_{ZZ'}}{m^2_{Z'}},\\
m^2_{Z_2} &=& \fr 1 2 \left[m^2_{Z}+m^2_{Z'}+\sqrt{(m^2_Z-m^2_{Z'})^2+4m^4_{ZZ'}}\right] \simeq m^2_{Z'}.\eea The $Z$-$Z'$ mixing, i.e. $\varphi$, is small as suppressed by $(v_1,v_2)^2/\La^2$. The field $Z_1$ has a mass approximating that of $Z$, called the standard model $Z$-like boson, whereas the  field $Z_2$ is a new heavy gauge boson with mass proportional to $\La$. 

\section{$W$-boson mass deviation}

The model under consideration yields a tree-level mixing of $Z$ with $Z'$. Because of this mixing, the observed $Z_1$ mass is reduced, compared with the standard model $Z$ mass. Additionally, this reduction gives rise to a positive contribution to $T$-parameter, 
\be \al T=\rho-1=\fr{m^2_W}{c^2_W m^2_{Z_1}}-1\simeq \fr{m^4_{ZZ'}}{m^2_Z m^2_{Z'}}\simeq \fr{\left[s^2_{\theta} v^2_1+(s^2_\theta+4x)v^2_2\right]^2}{16x^2(v^2_1+v^2_2)\La^2},\ee with the aid of $m_W=m_Zc_W$ and $m_{Z_1}$ approximated above. Since the $Z_1$ mass is fixed as precisely measured, the positive value of $\al T$ dominantly enhances the $W$ mass (cf. \cite{stu,strumia}), \be \Delta m^2_{W}=\fr{c^4_W m^2_{Z}}{c^2_W-s^2_W}\al T \simeq \fr{g^2c^2_W\left[s^2_{\theta} v^2_1+(s^2_\theta+4x)v^2_2\right]^2}{64x^2 c_{2W}\La^2}.\ee 

\begin{figure}[h]
\bc
\includegraphics[scale=0.8]{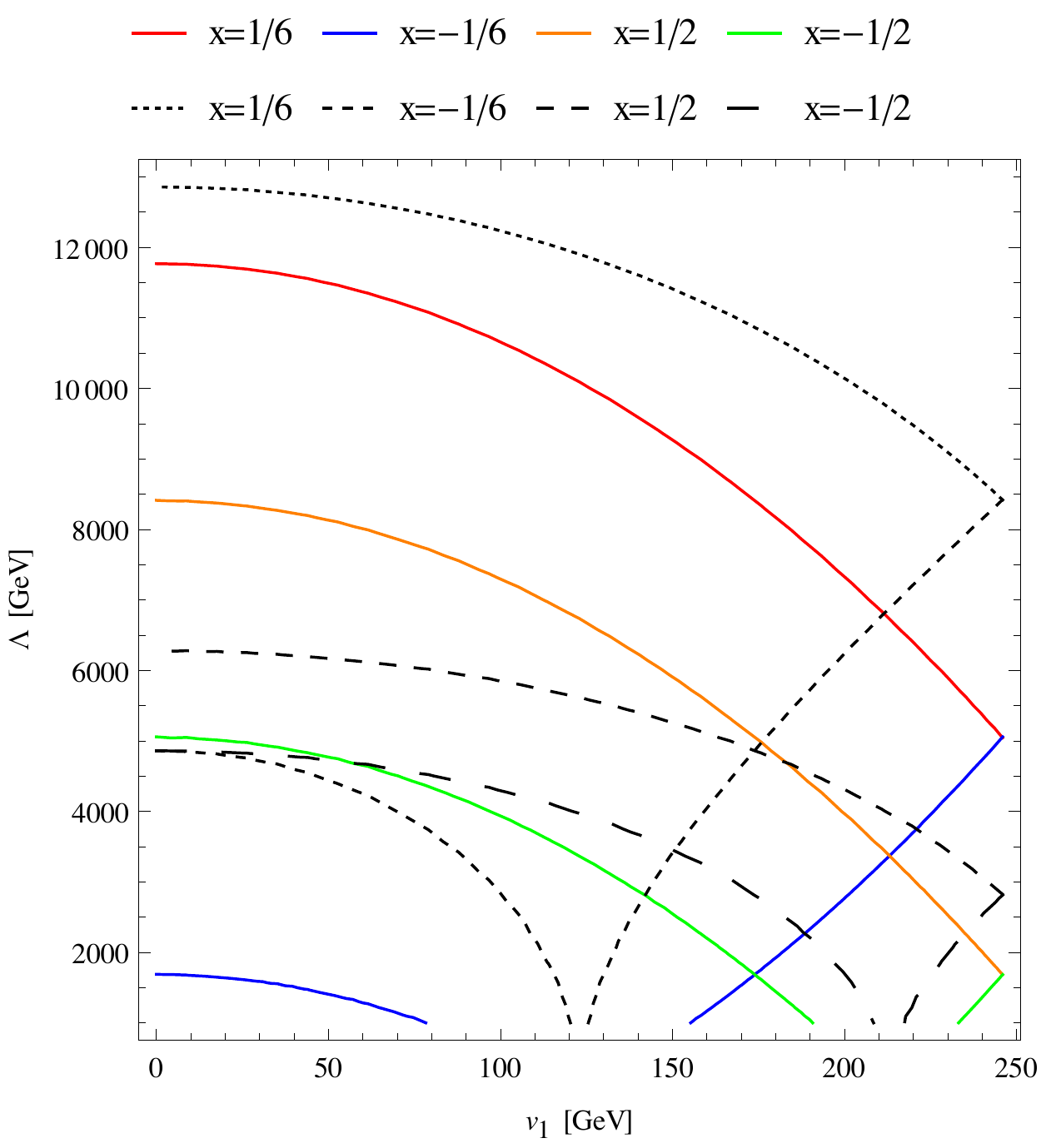}
\caption[]{\label{fig1} CDF $W$-mass deviation contoured as function of $(v_1,\La)$ according to several values of $x$-charge parameter (solid, color curves). New physics scale $\La$ limited by the electroweak precision test on $|\varphi|\lesssim 10^{-3}$ as correspondingly combined (dashed curves).}
\ec
\end{figure}

Use the recent measurement $m_W|_{\mathrm{CDF}}=80.4335 \pm 0.0094$ GeV \cite{cdf}, which deviates from the standard model prediction, $m_W|_{\mathrm{SM}}=80.357 \pm 0.006$ GeV, at 7$\sigma$. We make a contour of $\Delta m^2_W=m^2_W|_\mathrm{CDF}-m^2_{W}|_{\mathrm{SM}}$ as function of $v_1$ and $\La$ for a choice of $x=\pm \fr 1 6$ and $\pm \fr 1 2 $, as in Figure \ref{fig1} (solid, color curves). Here we have used $v_2=\sqrt{(246\ \mathrm{GeV})^2-v^2_1}$, fixed $t_\theta=g_N/g_X=1$, and taken $s^2_W=0.231$ and $\al=1/128$. It is clear from the figure that concerning the $W$-mass deviation, the new physics scale $\La$ depends on $x$ for each $v_1$. For each $|x|$-value, $\La$ is larger for positive $x$, while $\La$ is smaller for negative $x$. Additionally, $\La$ is appropriate to the seesaw setup as well as the FCNC and collider bounds (discussed below) for interested values of $x$ with an appropriate choice of $v_1$. However, the scale $\La$ is being constrained by $Z$-fermion couplings, specified in the following section.       

\section{Interactions}

Substituting the gauge fields in terms of the physical fields obtained above, the covariant derivative takes the form, 
\bea D_\mu &=& \pa_\mu + ig_s t_p G_{p\mu} + i e Q A_\mu + i g (T_+W^+_\mu + H.c.)  \crn
&&+ \fr{ig}{c_W}\left[c_\varphi (T_3-s^2_W Q)-\fr{s_\varphi s_W}{s_\theta c_\theta}(X-s^2_\theta Y)\right]Z_{1\mu} \crn
&&+ \fr{ig}{c_W}\left[s_\varphi (T_3-s^2_W Q)+\fr{c_\varphi s_W}{s_\theta c_\theta}(X-s^2_\theta Y)\right]Z_{2\mu},\eea where $T_\pm=(T_1\pm i T_2)/\sqrt{2}$. Substituting this covariant derivative into the fermion kinetic term, $\sum_F \bar{F}i\ga^\mu D_\mu F$, the interactions of gauge bosons with fermions are derived. 

It is clear that gluons, photon, and $W^\pm$ interact with fermions as in the standard model,
\bea \mathcal{L}_{\mathrm{int}} &\supset& -\fr{g_s}{2} \bar{q} \ga^\mu \la_p q G_{p\mu}-eQ(f)\bar{f}\ga^\mu f A_\mu - \fr{g}{\sqrt{2}} [(\bar{\nu}_L \ga^\mu e_L + \bar{u}_L\ga^\mu d_L)W^+_\mu+H.c.],\eea where $\la_p$ for $p=1,2,\cdots,8$ denotes the Gell-Mann matrices. Additionally, $q$ and $f$ indicate every quark and fermion in mass eigenstate basis, while $\nu$, $e$, $u$, and $d$ stand for neutrinos, charged leptons, up quarks, and down quarks in gauge basis, respectively. 

The interactions of $Z_{1,2}$ with fermions have the form,
\be \mathcal{L}_{\mathrm{int}}\supset -\fr{g}{2c_W}\bar{f}\ga^\mu [g^{Z_I}_V(f)- g^{Z_I}_A(f)\ga_5]f Z_{I\mu},\ee where $I=1,2$, and 
\bea g^{Z_1}_V(f) &=& (c_\varphi-s_\varphi s_W t_\theta) T_3(f_L)\crn
&&-2s_W(c_\varphi s_W-s_\varphi t_\theta)Q(f)-2(s_\varphi s_W/s_\theta c_\theta)X(f),\\
g^{Z_1}_A(f) &=&(c_\varphi -s_\varphi s_W t_\theta)T_3(f_L),\\
g^{Z_2}_{V,A}(f) &=& g^{Z_1}_{V,A}(f)|_{c_\varphi\rightarrow s_\varphi, s_\varphi\rightarrow -c_\varphi}.\eea The $V,A$ couplings of $Z_{1,2}$ with fermions are collected in Table \ref{tab2} and \ref{tab3}, respectively.

\begin{table}[h]
\bc
\begin{tabular}{c|cc}
\hline\hline
$f$ & $g^{Z_1}_V(f)$ & $g^{Z_1}_A(f)$ \\
\hline 
$\nu_a$ & $\fr{c_\varphi}{2}-s_W s_\varphi\left(\fr{t_\theta}{2}+\fr{2x}{s_\theta c_\theta}\right)$ & $\fr 1 2\left(c_\varphi-s_W s_\varphi t_\theta\right)$\\
$e_a$ & $\fr{c_\varphi}{2}\left(4s^2_W-1\right)-s_W s_\varphi\left(\fr{3t_\theta}{2}+\fr{2x}{s_\theta c_\theta}\right)$ & $\fr 1 2\left(s_W s_\varphi t_\theta - c_\varphi\right)$\\
$u_\al$ & $\fr{c_\varphi}{6}\left(3-8s^2_W\right)+s_W s_\varphi\left(\fr{5t_\theta}{6}+\fr{2x}{s_\theta c_\theta}\right)$ & $\fr 1 2\left(c_\varphi-s_W s_\varphi t_\theta\right)$\\
$u_3$ & $\fr{c_\varphi}{6}\left(3-8s^2_W\right)+s_W s_\varphi\left(\fr{5t_\theta}{6}-\fr{2x}{s_\theta c_\theta}\right)$ & $\fr 1 2\left(c_\varphi-s_W s_\varphi t_\theta\right)$\\
$d_\al$ & $\fr{c_\varphi}{6}\left(4s^2_W-3\right)-s_W s_\varphi\left(\fr{t_\theta}{6}-\fr{2x}{s_\theta c_\theta}\right)$ & $\fr 1 2\left(s_W s_\varphi t_\theta - c_\varphi\right)$\\
$d_3$ & $\fr{c_\varphi}{6}\left(4s^2_W-3\right)-s_W s_\varphi\left(\fr{t_\theta}{6}+\fr{2x}{s_\theta c_\theta}\right)$ & $\fr 1 2\left(s_W s_\varphi t_\theta - c_\varphi\right)$\\
\hline\hline
\end{tabular}
\caption[]{\label{tab2} Couplings of $Z_1$ with fermions.}
\ec
\end{table}

\begin{table}[h]
\bc
\begin{tabular}{c|cc}
\hline\hline
$f$ & $g^{Z_2}_V(f)$ & $g^{Z_2}_A(f)$ \\
\hline 
$\nu_a$ & $\fr{s_\varphi}{2}+s_W c_\varphi\left(\fr{t_\theta}{2}+\fr{2x}{s_\theta c_\theta}\right)$ & $\fr 1 2\left(s_\varphi+s_W c_\varphi t_\theta\right)$\\
$e_a$ & $\fr{s_\varphi}{2}\left(4s^2_W-1\right)+s_W c_\varphi\left(\fr{3t_\theta}{2}+\fr{2x}{s_\theta c_\theta}\right)$ & $-\fr 1 2\left(s_W c_\varphi t_\theta + s_\varphi\right)$\\
$u_\al$ & $\fr{s_\varphi}{6}\left(3-8s^2_W\right)-s_W c_\varphi\left(\fr{5t_\theta}{6}+\fr{2x}{s_\theta c_\theta}\right)$ & $\fr 1 2\left(s_\varphi+s_W c_\varphi t_\theta\right)$\\
$u_3$ & $\fr{s_\varphi}{6}\left(3-8s^2_W\right)-s_W c_\varphi\left(\fr{5t_\theta}{6}-\fr{2x}{s_\theta c_\theta}\right)$ & $\fr 1 2\left(s_\varphi+s_W c_\varphi t_\theta\right)$\\
$d_\al$ & $\fr{s_\varphi}{6}\left(4s^2_W-3\right)+s_W c_\varphi\left(\fr{t_\theta}{6}-\fr{2x}{s_\theta c_\theta}\right)$ & $-\fr 1 2\left(s_W c_\varphi t_\theta + s_\varphi\right)$\\
$d_3$ & $\fr{s_\varphi}{6}\left(4s^2_W-3\right)+s_W c_\varphi\left(\fr{t_\theta}{6}+\fr{2x}{s_\theta c_\theta}\right)$ & $-\fr 1 2\left(s_W c_\varphi t_\theta + s_\varphi\right)$\\
\hline\hline
\end{tabular}
\caption[]{\label{tab3} Couplings of $Z_2$ with fermions.}
\ec
\end{table}

Notice that $|\varphi|\ll 1$ as substantially suppressed by $(v_1,v_2)^2/\La^2$. In the limit $\varphi=0$, the field $Z_1$ becomes the standard model $Z$ boson, and the $V,A$ couplings of $Z_1$ become those of $Z$ in the standard model, i.e. $g^{Z_1}_{V,A}(f)\rightarrow g^{Z}_{V,A}(f)$ for $\varphi\rightarrow 0$. In other words, the current theory is consistent with the standard model at low energy. When $0\neq |\varphi|\ll 1$, the well-measured couplings of $Z$ with fermions are modified by an amount proportional to $\varphi$, 
\bea g^{Z_1}_V(f) &=& T_3(f_L)-2s^2_W Q(f)+ \mathcal{O}(\varphi), \\ g^{Z_1}_A(f) &=& T_3(f_L) + \mathcal{O}(\varphi),\eea where note that $\theta$ is always finite. With the aid of the electroweak precision data \cite{pdg}, the new physics contribution is safe, given that $|\varphi|\lesssim 10^{-3}$ which applies for most $Z'$ models including ours. Indeed, the electroweak precision data include low-energy weak neutral current experiments, precision $Z$-pole physics, and others. The low-energy weak neutral current experiments are modified by $Z'$ exchange, mainly sensitive to its mass. The precision $Z$-pole physics at the LEP and SLC are mainly sensitive to $Z$-$Z'$ mixing, which lowers the $Z$ mass and shifts the $Z$-fermion couplings, by contrast. Notice that the $Z$-fermion couplings will determine the various $Z$-pole observables, such as asymmetries and partial widths. The precision data strongly constrain the $Z$-$Z'$ mixing angle, $\varphi$, which is extracted by comparing the very precise measurements of the coupling constants at the $Z$ pole with the standard model predictions including radiative corrections (cf. \cite{zzprimemixing}). They also provide lower limits on the $Z'$ mass, but these limits are generally weaker than the LEPII and LHC, obtained later (these high-energy experiments that are above the $Z$-pole will supply strong constraints on the $Z'$ mass, but generally not sensitive to small $Z$-$Z'$ mixing). The precision data constraint can be presented for two cases of $\rho_0$-paramter, $\rho_0$ free for arbitrary Higgs structure and $\rho_0=1$ for Higgs doublets and singlets like ours, but there is only a slight difference between the limits given. Alternatively, performing a global fit to the full electroweak data set necessarily gives comparable bounds as $|\varphi|\sim 10^{-3}$, see \cite{lepewwg} for instance.   

With the choice of parameters as given above, the limit $|\varphi|\lesssim 10^{-3}$ produces a lower bound on $\La$ for each $v_1$, as despited in Figure \ref{fig1} (dashed curves) for a comparison to $W$ mass (solid, color curves). Only the dashed curves with $x=\pm1/2$ are suitable to the $W$-mass deviation (solid, color curves with $x=\pm1/2$) for $v_1<185$ GeV and $v_1<55$ GeV, respectively. By contrast, the dashed curves with $x=\pm 1/6$ are significantly bigger than those from the $W$-mass constraint, thus ruling out the values $x=\pm 1/6$ for every $v_1$. That said, the viable parameter space can be considered as ($x=1/2$, $\La=4.5$--8.5 TeV, $v_1=0$--185 GeV) and ($x=-1/2$, $\La\sim 5$ TeV, $v_1=0$--55 GeV).    

Last, but not least, in contrast to the standard model, both $Z_{1,2}$ flavor-change when interacting with quarks, while conserving lepton flavors. This is due to the fact that quark families transform differently under $X,N$, while lepton families transform the same under such charges. Additionally, quark families transform the same under $T_3$ and $Y$ that define $Z$. Hence, the flavor changing effect associated with $Z_1$ results from a $Z$-$Z'$ mixing to be small. Whereas, the flavor change dominantly occurs associated with $Z_2$, even for $\varphi=0$. We will investigate the FCNCs coupled to $Z_2$, while those with $Z_1$ are safe, given that $|\varphi|\lesssim 10^{-3}$.

\section{FCNCs}

The couplings of $Z_{1,2}$ with fermions arise from 
\be \mathcal{L}_{\mathrm{int}}\supset -\sum_F \bar{F}\ga^\mu (gT_3 A_{3\mu} + g_X X B_\mu + g_N N C_\mu) F.\ee Changing to the basis of $(A,Z,Z')$, i.e.  \bea A_3 &=& s_W A + c_W Z,\\ 
B &=& c_W s_\theta A - s_W s_\theta Z + c_\theta Z',\\ 
C &=& c_W c_\theta A - s_W c_\theta Z -s_\theta Z',\eea and by using $Q=T_3+Y$, $Y=X+N$, $e=gs_W$, and $g_X s_\theta=g_N c_\theta=g t_W$, we obtain  
\be \mathcal{L}_{\mathrm{int}}\supset -\sum_F \bar{F}\ga^\mu \left[e Q A_\mu +\fr{g}{c_W}(T_3-s^2_W Q)Z_\mu +\fr{g t_W}{s_\theta c_\theta} (X-s^2_\theta Y)Z'_\mu \right] F.\ee
Because $Q$, $T_3$, and $Y$ are universal for every repeated fermion flavors, such as neutrinos, charged leptons, up-type quarks, down-type quarks, the flavor change is only associated with $X$-charge. Additionally, lepton flavors are universal under $X$. Hence, this flavor change occurs only for quarks. The relevant Lagrangian is 
\be \mathcal{L}_{\mathrm{int}} \supset -\fr{gt_W}{s_\theta c_\theta } \bar{q}_a\ga^\mu X q_a Z'_\mu 
= x \fr{gt_W}{s_\theta c_\theta } (\bar{q}_1\ga^\mu q_1 +\bar{q}_2 \ga^\mu q_2 - \bar{q}_3\ga^\mu q_3)Z'_\mu,\label{ptdt}\ee where $q_a$ denotes quarks of either up-types $(u_\al, u_3)$ or down-types $(d_\al, d_3)$, and notice also that $q_a$ is vector-like under $X$.  

Changing to the mass basis,
\bea q_{aL}=[V_{qL}]_{ai}q_{iL},\hs q_{aR}=[V_{qR}]_{ai} q_{iR},\eea where $q=u,d$, and $i=1,2,3$ labels mass eigenstates, $u_i=(u,c,t)$ and $d_i=(d,s,b)$, so that the quark mass matrices $m_{u,d}$ are diagonalized, 
\be V^\dagger_{uL} m_u V_{uR} = \mathrm{diag}(m_u,m_c,m_t),\hs V^\dagger_{dL} m_d V_{dR} = \mathrm{diag}(m_d,m_s,m_b). \ee Here, note that the CKM matrix is $V_{\mathrm{CKM}}=V^\dagger_{uL}V_{dL}$. Additionally, the quark current in (\ref{ptdt}) couples only the same quark chirality, i.e. $q\ga^\mu q =q_L \ga^\mu q_L + q_R \ga^\mu q_R$. Using the unitarity condition, $V^\dagger_{qL} V_{qL}=1=V^\dagger_{qR}V_{qR}$, we rewrite (\ref{ptdt}) as
\be \mathcal{L}_{\mathrm{int}} \supset  x \fr{gt_W}{s_\theta c_\theta } \bar{q}_i \ga^\mu q_i Z'_\mu  -2 x \fr{gt_W}{s_\theta c_\theta } [V^*_{qL}]_{3i}[V_{qL}]_{3j} \bar{q}_{iL}\ga^\mu q_{jL}Z'_\mu + (L\rightarrow R),\ee where $i,j$ are summed. The first term conserves quark flavors, whereas the last two terms give rise to FCNCs for $i\neq j$,
\be \mathcal{L}_{\mathrm{FCNC}}=L_{ij} \bar{q}_{iL}\ga^\mu q_{jL} Z'_\mu + (L\rightarrow R),\ee where we have defined,
\be L_{ij}=-2x\fr{gt_W}{s_\theta c_\theta } [V^*_{qL}]_{3i}[V_{qL}]_{3j}.\ee  

Substituting $Z'=c_\varphi Z_2-s_\varphi Z_1$ into the FCNC Lagrangian and integrating $Z_{1,2}$ out, we get effective interactions,
\be \mathcal{H}^{\mathrm{eff}}_{\mathrm{FCNC}}= \fr{L^2_{ij}}{m^2_{Z_2}} (\bar{q}_{iL}\ga^\mu q_{jL})^2+ 2(LR)+(RR),\label{ptdt11}\ee where the last two terms differ from the first one only in chiral structures. We have neglected the contribution of $Z_1$, because it is more suppressed than that of $Z_2$, i.e. \be 
\fr{s^2_\varphi}{m^2_{Z_1}} \ll \fr{c^2_\varphi}{m^2_{Z_2}},\ee due to $t_\varphi \sim m^2_{Z_1}/m^2_{Z_2}$; additionally, we have utilized $c_\varphi\simeq 1$. The effective couplings in (\ref{ptdt11}) are further approximated as \be \fr{L^2_{ij}}{m^2_{Z_2}}\simeq \fr{1}{\La^2}([V^*_{qL}]_{3i}[V_{qL}]_{3j})^2,\label{ptdt12}\ee and similar for the others, which are all independent of $\theta,x$, since $\La\gg v_1,v_2$. 

The effective interactions in (\ref{ptdt11}) contribute to the neutral-meson mixing amplitudes, extensively studied in the literature. Assuming the dominant new physics effects come from the first term, $L^2_{ij}$, the existing data imply 
\be \fr{L^2_{12}}{m^2_{Z_2}}<\left(\fr{1}{10^4\ \mathrm{TeV}}\right)^2,\hs \fr{L^2_{13}}{m^2_{Z_2}}<\left(\fr{1}{500\ \mathrm{TeV}}\right)^2,\hs \fr{L^2_{23}}{m^2_{Z_2}}<\left(\fr{1}{100\ \mathrm{TeV}}\right)^2,\label{ptdt15}\ee according to $K^0$-$\bar{K}^0$ mixing, $B^0_d$-$\bar{B}^0_d$ mixing, and $B^0_s$-$\bar{B}^0_s$ mixing, respectively \cite{fcnceff}. Aligning the quark mixing to the down quark sector, without loss of generality, we have $V_{uL}=1$, thus $V_{\mathrm{CKM}}=V_{dL}$. Since $V_{\mathrm{CKM}}$ is well measured, we input \cite{pdg} 
\be [V_{dL}]_{31}=0.00886,\hs [V_{dL}]_{32}=0.0405,\hs [V_{dL}]_{33}=0.99914.\ee Hence, the constraints in (\ref{ptdt15}) yield
\be \La> 3.6\ \mathrm{TeV},\hs \La>4.4\ \mathrm{TeV},\hs \La>4\ \mathrm{TeV},\label{ptdt112}\ee corresponding to the mentioned meson mixing systems, compatible to the previous bounds.  

Alternatively, the effective interactions $(LR)$ and $(RR)$ also contribute to the neutral-meson mixing amplitudes through switching on the right-handed quark mixing matrix, $V_{qR}$. Since $V_{qR}$ is left arbitrarily as in the standard model (where only the left-handed quark mixing is constrained by the CKM matrix $V_{\mathrm{CKM}}=V^\dagger_{uL}V_{dL}$), an issue arises that this $V_{qR}$ contribution leads to dangerous FCNCs? However, as given from outset, our theory originates from a trinification which conserves a left-right symmetry. We expect that the left-right symmetry is still good at the energy scale of the current theory, i.e. $V_{qR}\sim V_{qL}$. This would suppress the dangerous $(LR)$ and $(RR)$ contributions to neutral-meson mixings as the CKM factor does so for the $(LL)$ coupling. To be concrete, we first consider the contribution of all effective interactions in (\ref{ptdt11}) to the mass difference of $K^0$-$\bar{K}^0$ mixing system, such as  
\be \Delta m_{K}=2\mathrm{Re}\langle K^0|\mathcal{H}^{\mathrm{eff}}_{\mathrm{FCNC}}|\bar{K}^0\rangle,\ee where $K^0$ is constructed from two quarks $(q_i,q_j)=(d,s)$. Further, we have 
\bea \Delta m_K &=& 2 \mathrm{Re}\langle K^0|\fr{L^2_{12}}{m^2_{Z_2}}(\bar{d}_L \ga^\mu s_L)^2+2\fr{L_{12}R_{12}}{m^2_{Z_2}}(\bar{d}_L \ga^\mu s_L)(\bar{d}_R \ga_\mu s_R)+\fr{R^2_{12}}{m^2_{Z_2}}(\bar{d}_R \ga^\mu s_R)^2|\bar{K}^0\rangle \crn
&\simeq& \fr{2m_K f^2_K}{3m^2_{Z_2}} \mathrm{Re}\left\{L^2_{12}-\left[\fr 3 2+ \left(\fr{m_K}{m_d+m_s}\right)^2\right]L_{12}R_{12}+R^2_{12}\right\},\label{ptdt16}\eea with the aid of 
\bea && \langle K^0|(\bar{d}_L \ga^\mu s_L)^2|\bar{K}^0\rangle= \langle K^0|(\bar{d}_R \ga^\mu s_R)^2|\bar{K}^0\rangle=\fr 1 3 m_K f^2_K,\\ 
&& \langle K^0|(\bar{d}_L \ga^\mu s_L)(\bar{d}_R \ga_\mu s_R)|\bar{K}^0\rangle = -\fr 1 2 \left[\fr 1 2 +\fr 1 3 \left(\fr{m_K}{m_d+m_s}\right)^2\right] m_K f^2_K,\eea in agreement with \cite{massdiff}. Here, the hadronic matrix elements have been determined by the vacuum insertion approximation using PCAC \cite{mm}. Similarly for $B^0_{d,s}$-$\bar{B}^0_{d,s}$ mixings, i.e. $(q_i,q_j)=(d,b)$ and $(s,b)$, respectively, we have 
\bea \Delta m_{B_d} &\simeq& \fr{2m_{B_d} f^2_{B_d}}{3m^2_{Z_2}} \mathrm{Re}\left\{ L^2_{13}-\left[\fr 3 2+ \left(\fr{m_{B_d}}{m_d+m_b}\right)^2\right]L_{13}R_{13}+R^2_{13}\right\},\label{ptdt17}\\ 
\Delta m_{B_s} &\simeq& \fr{2m_{B_s} f^2_{B_s}}{3m^2_{Z_2}} \mathrm{Re}\left\{ L^2_{23}-\left[\fr 3 2+ \left(\fr{m_{B_s}}{m_s+m_b}\right)^2\right]L_{23}R_{23}+R^2_{23}\right\}.\label{ptdt18}\eea The contribution terms of new physics in (\ref{ptdt16}), (\ref{ptdt17}), and (\ref{ptdt18}) are equally suppressed by an approximate left-right symmetry, $V_{dR}\sim V_{dL}$, which implies $R_{ij}\sim L_{ij}$. Assuming the contributions $(LR)$ and $(RR)$ equivalent to $(LL)$, using the approximation (\ref{ptdt12}), the $\La$ bounds from the mass differences (cf. \cite{massdiff}) are compatible to the ones determined in (\ref{ptdt112}). 

\section{Collider bounds}

Since $Z$ and $Z'$ slightly mix, we will omit their mixing effect in this section, without loss of generality. That said, we set $\varphi\to 0$ and $Z_2\to Z'$ as a physical field. Because $Z'$ interacts with both usual quarks and leptons, it definitely presents interesting signals at the current hadron and lepton colliders. 

The LEPII experiment \cite{lep2} has studied the process $e^+e^- \to f\bar{f}$ that produces a pair of ordinary fermions $(f\bar{f})$ through exchange of $Z'$. This process is best described by the following effective interactions,
\be \mathcal{L}^{\mathrm{eff}}_{\mathrm{LEP2}}\supset \fr{g^2}{c^2_Wm^2_{Z'}}\left[\bar{e}\ga^\mu(a^{Z'}_L(e)P_L+a^{Z'}_R(e)P_R)e\right]\left[\bar{f}\ga_\mu(a^{Z'}_L(f)P_L+a^{Z'}_R(f)P_R)f\right],\ee where the chiral gauge couplings $a^{Z'}_{L,R}(f)=\fr 1 2 [g^{Z'}_V(f)\pm g^{Z'}_A(f)]$, as usual. Considering dilepton signals, $f=\mu,\tau$, all the leptons possess universal couplings, such that 
\be \mathcal{L}^{\mathrm{eff}}_{\mathrm{LEPII}}\supset \fr{g^2[a^{Z'}_L(e)]^2}{c^2_W m^2_{Z'}}(\bar{e}\ga^\mu P_L e)(\bar{f}\ga_\mu P_Lf)+(LR)+(RL)+(RR),\ee where the explicit couplings are given, using Table \ref{tab3},   
\be a^{Z'}_L (e) = \fr{s_W(s^2_\theta + 2x)}{s_{2\theta}},\hs a^{Z'}_R(e)=\fr{s_W(s^2_\theta+x)}{s_\theta c_\theta}. \ee 

The LEPII experiment has searched for such chiral interactions, giving several constraints on respective couplings, commonly achieved at order of a few TeV. Considering the bound derived for a new $U(1)$ gauge boson like ours, it typically obeys \cite{carena}  
\be \fr{g^2[a^{Z'}_L(e)]^2}{c^2_W m^2_{Z'}} <\fr{1}{(6\ \mathrm{TeV})^2}.\ee Using the approximation, $m^2_{Z'}\simeq (16x^2g^2t^2_W/s^2_{2\theta})\La^2$, we get 
\be \La> 1.5 (2+s^2_\theta/|x|)\ \mathrm{TeV}=4.5\ \mathrm{TeV},\ee for $t_\theta =1$ and $|x|=1/2$, as before, in agreement with the previous bounds.   

\begin{figure}[h]
\bc
\includegraphics[scale=0.7]{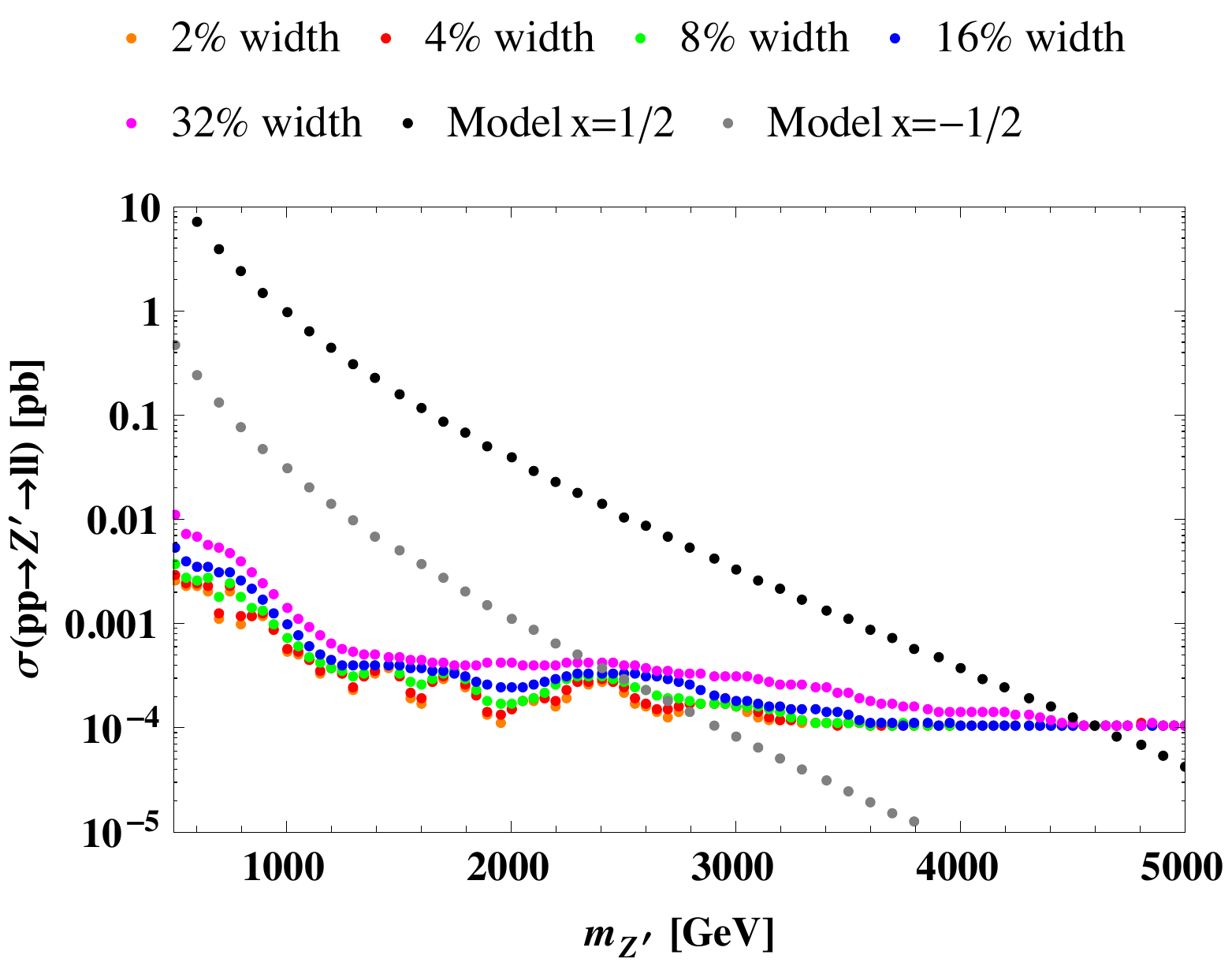}
\caption{\label{fig2} Dilepton production cross-section as function of $Z'$ mass, where dotted lines denote observed limit for different widths extracted at dilepton invariant-mass resonance, using 36.1 fb$^{-1}$ of $pp$ collision at $\sqrt{s}=13$ TeV by ATLAS \cite{lhc}. The black and gray dotted lines are theoretical prediction according to $x=1/2$ and $-1/2$, respectively.}
\ec
\end{figure}

At the LHC experiment, both new physics signals of dijet \cite{dijet} and dilepton \cite{lhc} have been investigated. However, since the current bound for dijet is less sensitive than dilepton, we will only examine the latter. The cross-section producing a final state of dilepton can be estimated, using narrow width approximation, such as
\be \sigma(pp\to Z'\to f \bar{f})=\fr 1 3 \sum_q \fr{dL_{q\bar{q}}}{dm^2_{Z'}}\hat{\sigma}(q\bar{q}\to Z')\mathrm{Br}(Z'\to f\bar{f}),  \ee  
where the luminosity $dL_{q\bar{q}}/dm^2_{Z'}$ can be extracted from \cite{lumi} for LHC $\sqrt{s}=13$ TeV. The partonic cross-section and branching $\mathrm{Br}(Z'\to f\bar{f})=\Ga(Z'\to f\bar{f})/\Ga_{Z'}$ take the form, 
\bea
&&\hat{\sigma}(q\bar{q}\to Z')=\fr{\pi g^2}{12c^2_W}[(g^{Z'}_V(q))^2+(g^{Z'}_A(q))^2],\\
&& \Ga(Z'\to f\bar{f})=\fr{g^2m_{Z'}}{48\pi c^2_W}N_C(f)[(g^{Z'}_V(f))^2+(g^{Z'}_A(f))^2],\\
&& \Ga_{Z'}=\fr{g^2m_{Z'}}{48\pi c^2_W}\sum_{f} N_C(f) [(g^{Z'}_V(f))^2+(g^{Z'}_A(f))^2],\eea where we exclude the decay $Z'\to \nu_R\bar{\nu}_R$, and note also that $Z'$ decay to other fields negligibly contributes to $Z'$ width. In Figure \ref{fig2}, we plot the dilepton production cross-section for $f=e,\mu,\tau$, commonly labelled $l$, which have the same $Z'$ couplings, thus cross-section. The ATLAS \cite{lhc} yields a negative result for new dilepton event of high mass, thus this translates to a lower limit for $Z'$ mass $m_{Z'}> 4.6$ TeV and $m_{Z'}>2.8$ TeV for the model with $x=1/2$ and $x=-1/2$, respectively, where $t_\theta=1$ is taken as above. Hence, using $m_{Z'}\simeq (4 |x| g t_W/s_{2\theta})\La $ leads to $\La>6.4$ TeV and $\La>3.9$ TeV according to the $Z'$ mass bounds. These LHC bounds are in agreement with the previous constraints.   

\section{Conclusion}

A guide from potential grand unification is that fermion families at low energy might be not universal, unlike the conventional standard model. This work has shown that such abelian remnants are presented, making a match of family number to color number. Additionally, the hypercharge is a result of their breaking, making an electrically-neutral world whose proton and electron charges are cancelled out. 

Interestingly, the Higgs fields experience the new abelian charges, causing a mixing of the usual $Z$ boson with new $Z'$ boson, explaining the recently-measured $W$-boson mass anomaly. The electroweak precision data on $Z$-$Z'$ mixing combined with the $W$ mass measurement implies the new physics scale $\La=4.5$--8.5 TeV and a weak scale $v_1=0$--185 GeV for the charge parameter $x=1/2$, while $\La\sim 5$ TeV and $v_1=0$--55 GeV for $x=-1/2$, where for all cases, the relative new gauge couplings set as $t_\theta=g_N/g_X=1$. Additionally, the values of $x=\pm 1/6$ are obviously ruled out. 

This theory contains the right-handed neutrinos as fundamental constituents, and the symmetry breaking leads to naturally-small neutrino masses, suppressed by the new physics scale that sets Majorana right-handed neutrino masses also at TeV scale.     

The family nonuniversality occurs only for quarks, and this leads to FCNCs of quarks coupled to $Z'$. An analysis for the contribution of left currents $(LL)$ implies a bound for new physics scale $\La>$ 3.6 TeV, 4 TeV, and 4.4 TeV according to the mixings, $K^0$-$\bar{K}^0$, $B^0_{s}$-$\bar{B}^0_{s}$, and $B^0_{d}$-$\bar{B}^0_{d}$, respectively. The contribution of right currents such as $(LR)$ and $(RR)$ to the mixings may be large, but safely suppressed by assuming an approximate left-right symmetry. This kind of constraint is independent of $x,\theta$. 

Considering $Z'$ contribution to dilepton signals at LEPII gives a bound $\La>4.5$ TeV for $|x|=1/2$ and $t_\theta=1$, while the dilepton production cross-section at LHC induced by $Z'$ implies $\La>$ 6.4 TeV and 3.9 TeV with respect to $x=1/2$ and $x=-1/2$, respectively, while keeping $t_\theta=1$. Hence, the collider, FCNC, electroweak precision data, and $W$-mass measurement on the new physics scale are all compatible.                  

If the current theory is embedded in a trinification \cite{trinification} or $E_6$ \cite{e6}, this implies $x=\pm 1/6$, as well as the family nonuniversality is lost. An analysis shows that our proposal can be properly embedded in a flipped trinification \cite{hd} for arbitrary $x$ as the smallest choice of a unification symmetry.

\end{document}